\documentclass[journal,twoside,web]{ieeecolor}
\usepackage{generic}
\usepackage{cite}
\usepackage{amsmath,amssymb,amsfonts}
\usepackage{algpseudocode}
\usepackage{algorithm}
\usepackage{array}
\newcolumntype{M}[1]{>{\centering\arraybackslash}m{#1}}
\newcolumntype{N}{@{}m{0pt}@{}}
\usepackage[switch,columnwise]{lineno}
\usepackage{hhline}
\usepackage{pifont}
\newcommand{\cmark}{\ding{51}}%
\newcommand{\xmark}{\ding{55}}%
\usepackage{graphicx}
\usepackage{textcomp}
\usepackage{siunitx}
\usepackage[bookmarks=false]{hyperref}
\def\BibTeX{{\rm B\kern-.05em{\sc i\kern-.025em b}\kern-.08em
    T\kern-.1667em\lower.7ex\hbox{E}\kern-.125emX}}
\definecolor{blue}{rgb}{0, 0, 1}

\usepackage{eso-pic}

\AddToShipoutPicture*{\AtPageUpperLeft{\scriptsize \sffamily\raisebox{-0.7cm}{\hspace{2.2cm}This article has been accepted for publication in IEEE Journal of Biomedical and Health Informatics licensed under a Creative Commons Attribution 4.0 License.}}}

\begin{document}
\title{Adaptive Body Area Networks Using Kinematics and Biosignals}
\author{Ali Moin, \IEEEmembership{Student Member, IEEE}, Arno Thielens, \IEEEmembership{Member, IEEE}, Alvaro Araujo, \IEEEmembership{Member, IEEE}, Alberto~Sangiovanni-Vincentelli,~\IEEEmembership{Fellow, IEEE}, and Jan M. Rabaey, \IEEEmembership{Fellow, IEEE}
\thanks{Manuscript received December 10, 2019. This work was supported in part by the CONIX Research Center, one of six centers in JUMP, a Semiconductor Research Corporation (SRC) program sponsored by DARPA. Support was also received from sponsors of Berkeley Wireless Research Center. \emph{(Corresponding author: Ali Moin.)}}
\thanks{A.\ Moin, A.\ Thielens, A.\ Sangiovanni-Vincentelli, and J.\ M.\ Rabaey are with the Department of Electrical Engineering and Computer Sciences, University of California at Berkeley, Berkeley, CA 94720 USA (e-mail: moin@berkeley.edu; arno.thielens@berkeley.edu; alberto@berkeley.edu; jan\_rabaey@berkeley.edu).}
\thanks{A. Araujo is with B105 Electronic Systems Lab, Universidad Politecnica de Madrid, Madrid, Spain (e-mail: araujo@b105.upm.es).}}

\maketitle

\begin{abstract}

The increasing penetration of wearable and implantable devices necessitates energy-efficient and robust ways of connecting them to each other and to the cloud. However, the wireless channel around the human body poses unique challenges such as a high and variable path-loss caused by frequent changes in the relative node positions as well as the surrounding environment. An adaptive wireless body area network (WBAN) scheme is presented that reconfigures the network by learning from body kinematics and biosignals. It has very low overhead since these signals are already captured by the WBAN sensor nodes to support their basic functionality. Periodic channel fluctuations in activities like walking can be exploited by reusing accelerometer data and scheduling packet transmissions at optimal times. Network states can be predicted based on changes in observed biosignals to reconfigure the network parameters in real time. A realistic body channel emulator that evaluates the path-loss for everyday human activities was developed to assess the efficacy of the proposed techniques. Simulation results show up to 41\% improvement in packet delivery ratio (PDR) and up to 27\% reduction in power consumption by intelligent scheduling at lower transmission power levels. Moreover, experimental results on a custom test-bed demonstrate an average PDR increase of 20\% and 18\% when using our adaptive EMG- and heart-rate-based transmission power control methods, respectively. The channel emulator and simulation code is made publicly available at \url{https://github.com/a-moin/wban-pathloss}.

\end{abstract}


\section{Introduction}
\label{sec:introduction}
\IEEEPARstart{W}{earable} and implantable devices are becoming more prevalent in people's everyday lives with applications ranging from health monitoring and rehabilitation to human augmentation and entertainment. These devices must communicate with each other and with the cloud as an integral part of the Internet of Things (IoT), highlighting the need for a network infrastructure around the human body, traditionally named \textit{wireless body area network} (WBAN)~\cite{cavallari2014survey} and more recently the \textit{Human Intranet}~\cite{rabaey2015human}.

\subsection{Motivation}

Energy efficiency and robustness are among the important qualifications for a well-designed WBAN. Extended battery lifetime directly affects network usability and user satisfaction, but energy resources are scarce due to node size and wearability constraints. Wireless link robustness is also crucial, especially in safety-critical applications such as drug delivery~\cite{patel2010applications} and control of prosthetic limbs~\cite{benatti2017prosthetic}. Achieving a reliable network while maintaining low power consumption is generally a challenging problem to address, particularly when dealing with the unique characteristics of the human body.

The wireless channel around the human body is not an ideal medium for radio frequency wave transmission~\cite{yazdandoost2009channel}. Lower frequencies require unacceptably large on-body antennas, whereas at higher frequencies the propagation loss quickly becomes large~\cite{ryckaert2004channel}. Additionally, the channel is highly dynamic due to fading and shadowing effects, mainly caused by changes in body posture and kinematics. In order to maintain the desired reliability, the transmission power at the transmitter must be high enough to reach the sensitivity level of the receiver at all times, including when the channel temporarily experiences high attenuation.
However, a constant high transmission power sacrifices overall energy efficiency and causes interference with other nodes in the same and neighboring networks. Overall, energy consumption caused by communication represents a substantial fraction of the total energy budget of the sensor node.

In order to solve this problem, transmission power control (TPC) protocols have been introduced that evaluate the channel state by monitoring the received signal strength indicator (RSSI) at the receiver and adapt the transmission power level at the transmitter accordingly~\cite{xiao2009transmission,kim2013link}. Although most transceivers have the RSSI data readily available at the receiver side with no extra overhead, additional control packets must be fed back from the receiver to the transmitter to readjust the transmission power. Therefore, these methods lead to extra traffic overhead, as well as latency in the response. This is even worse in the WBAN dynamic scenarios where the node locations are rapidly changing and frequent link-state packets are needed to keep it up-to-date, otherwise the acquired link-state information is outdated by the time of the next transmission. Finally in the case of periodic movements such as walking, scheduling packet transmissions at times that the channel has lower loss is a far better solution than blindly increasing Tx power and consequently worsening the interference~\cite{zang2017gait,roberts2012exploiting,prabh2012banmac}.

In this paper, we propose an adaptive network methodology that learns from body kinematics and biosignals, predicts the wireless channel behavior based on them, and subsequently reconfigures the network to achieve both \textit{energy efficiency} and \textit{robustness}. We measure the \textit{energy efficiency} based on the reduction in radio power consumption, and evaluate the \textit{robustness} in terms of the network packet delivery ratio (PDR) metric. Our method reuses the sensor data that already exists in a WBAN to adapt the network to different conditions with minimum overhead. These sensors include: 
\begin{itemize}
    \item inertial measurement units (IMU), which indicate body movements;
    \item electromyography (EMG) sensors measuring muscle activity (as currently integrated in prosthetic arms); 
    \item and heart rate (HR) monitoring sensors (extracted from electrocardiography [ECG] signals), which reveal increase in physical activity. 
\end{itemize}
The proposed adaptive schemes are specifically beneficial to WBANs as these generally suffer from the highly dynamic channel conditions and are severely constrained in terms of the energy and computation resources.

\subsection{Previous Work and Challenges}

A design space exploration method for WBAN was proposed in~\cite{moin2017optimized} to choose the best network topology and parameters for a given set of conditions before network deployment. In reality, however, there is no single best answer to this design problem due to the highly dynamic nature of the human body and its surroundings. Therefore, the network should be reconfigured adaptively based on its instantaneous state. Mechanisms such as TPC~\cite{xiao2009transmission,quwaider2010body,kim2013link,kim2013rssi} and link adaptation (e.g.\ changing modulation and data rate)~\cite{zhang2013energy} have been proven to be effective in the literature, as summarized in Table~\ref{tab:tpc}.

\begin{table*}[!t]

\caption{Summary of adaptive reconfiguration methods in WBANs.}
\label{tab:tpc}
\centering
\begin{tabular}{|M{3cm}||M{1cm}|M{1cm}|M{1.2cm}|M{1cm}|M{1.1cm}|M{1.4cm}|M{1cm}|M{1cm}|M{1.5cm}|N} \hline
 & Xiao \textit{et al.} \cite{xiao2009transmission} & Kim \& Eom \cite{kim2013link} & Quwaider \textit{et al.} \cite{quwaider2010body} & Kim \textit{et al.} \cite{kim2013rssi} & Zhang \textit{et al.} \cite{zhang2013energy} & Moulton \textit{et al.} \cite{moulton2010body} & Prabh \textit{et al.} \cite{prabh2012banmac} & Zang \textit{et al.} \cite{zang2017gait} & \textbf{This work} &\\[10pt] \hhline{|=#=|=|=|=|=|=|=|=|=|}
Constant control packets (from Rx to Tx) overhead & \cmark & \cmark & \cmark & \cmark & \cmark & \cmark (w/ adaptive periodicity) & \cmark & \cmark & \textbf{\xmark (only during IMU calibration)} &\\[10pt] \hline
Using RSSI data & \cmark & \cmark & \cmark & \cmark & \cmark & \cmark & \cmark & \cmark & \textbf{\cmark (during IMU calibration)} &\\[10pt] \hline
Using body kinematics & \xmark & \xmark & \xmark & \xmark & \xmark & \xmark & \cmark (from RSSI) & \cmark (from IMU) & \textbf{\cmark (from IMU)} &\\[10pt] \hline
Using bio-signals & \xmark & \xmark & \xmark & \xmark & \xmark & \xmark & \xmark & \xmark & \cmark &\\[10pt] \hline
\end{tabular}
\end{table*}

Most TPC methods are based on monitoring RSSI at the receiver and sending feedback to the transmitter~\cite{xiao2009transmission,quwaider2010body,kim2013link}. Kim \textit{et al.} in~\cite{kim2013rssi} argued that RSSI is not the most accurate indicator when packet failure is mainly due to interference instead of attenuation. In these cases, increasing the transmission power would even be harmful since it worsens interference. Moreover, all these methods suffer from the overhead of additional control packets from receiver to transmitter that increase their added energy consumption, traffic and latency that become then not negligible~\cite{moulton2010body}.

The dynamic wireless channel around the human body has some unique characteristics that can be leveraged. Roberts \textit{et al.}\ in~\cite{roberts2012exploiting} explore the periodicity of channel path-loss during certain movements and propose a new channel model, enhancing the widely used IEEE 802.15.6 model~\cite{yazdandoost2009channel}. Prabh \textit{et al.}\ in~\cite{prabh2012banmac} use this periodicity in RSSI to arrange packet transmissions at proper times. However, this approach has the same overhead problem since probing packets need to be broadcast to estimate RSSI and control packets need to be sent back to the transmitter. Zang \textit{et al.}\ in~\cite{zang2017gait} exploit periodic body movement patterns to schedule transmissions. They run a computationally-heavy template matching algorithm based on Dynamic Time Warping (DTW) to detect the strides. The entire algorithm has to run on the more powerful sink node and results on the next predicted transmission times are sent back to the transmitter. Hence, this method still faces the overhead problem of control packets.

In order to evaluate these methods, researchers use network simulators with human body channel models. Most of them are based on IEEE 802.15.6 standard~\cite{yazdandoost2009channel} that does not capture realistic activity-dependent fluctuations in path-loss due to body movements. Recently, Dautov and Tsouri proposed an off-body Rician channel model in~\cite{dautov2019dynamic} for indoor WBANs that addresses different types of body motions, but does not support on-body channels. Alam \textit{et al.}\ in~\cite{alam2016realistic} enhance the IEEE 802.15.6 on-body channel model by finding line-of-sight (LOS) and none-line-of-sight (NLOS) distances between nodes during real movements. They multiply the NLOS portion by a constant factor to partially account for shadowing effect. Alternatively, measurement campaigns using either vector network analyzers such as in~\cite{zasowski2003uwb,hu2007measurements,smith2009temporal} or simple radio nodes in~\cite{zang2017gait,quwaider2010body} can be used to estimate the channel loss in a WBAN. However, none of the nodes used in those studies have a biosignal acquisition front-end to enable the implementation of our proposed adaptive schemes.

\subsection{Our Contributions}

Our contributions can be summarized as follows: 1)~development of a body channel emulator capturing its dynamics during various movements. Use of this emulator in conjunction with our Castalia-based WBAN simulator~\cite{moin2017optimized,boulis2007castalia} enables realistic evaluation of the adaptive network framework; 2)~utilization of a custom-designed test-bed containing a \SI{2.4}{\giga\hertz} radio and biosignal front-ends to validate and evaluate the adaptive network schemes; 3)~measurement and analysis of on-body channel variations and stability during the main daily activities (standing, sitting, sleeping and walking); and 4)~development of computationally-light algorithms for adaptive network reconfiguration that includes: (i) a scheduling algorithm tracking periodic human movements without the constant overhead of control packets, (ii) an EMG-controlled TPC algorithm for reliable wireless links in prostheses, and (iii) an HR-controlled TPC scheme for enhanced robustness during body movements. These algorithms are simulated or implemented on the test-bed and experimental results are presented. To the best of our knowledge, this is the first time biosignals are reused for TPC in WBANs.

\section{Body Dynamics Emulator and test-bed}
In order to leverage the advantages of both simulation-based and empirical methods in our study, we developed (i) a realistic body channel emulator used to evaluate the performance of our proposed scheduling algorithm in realistic dynamic WBAN conditions; and (ii) a \SI{2.4}{\giga\hertz} wearable node with embedded neural front-ends, used to demonstrate the efficacy of our TPC methods in real scenarios.

\subsection{Human Body Dynamic Channel Emulator}
\label{sec:emulator}

The wireless channel around the human body is highly dynamic. Most WBAN simulators either assume a static path-loss between pairs of nodes or add a random variation factor to capture dynamic effects such as channel fading and shadowing~\cite{moin2017optimized,boulis2007castalia}. While these statistical models mimic the physical layer to some extent, they do not necessarily show the correlation between activity-specific kinematics of the human body and its path-loss. Hence, having a wireless channel emulator that generates the path-loss based on actual body movements is beneficial, especially when adaptive network protocols based on those dynamics are being tested.

\begin{figure*}[!t]
\centerline{\includegraphics[width=0.75\textwidth]{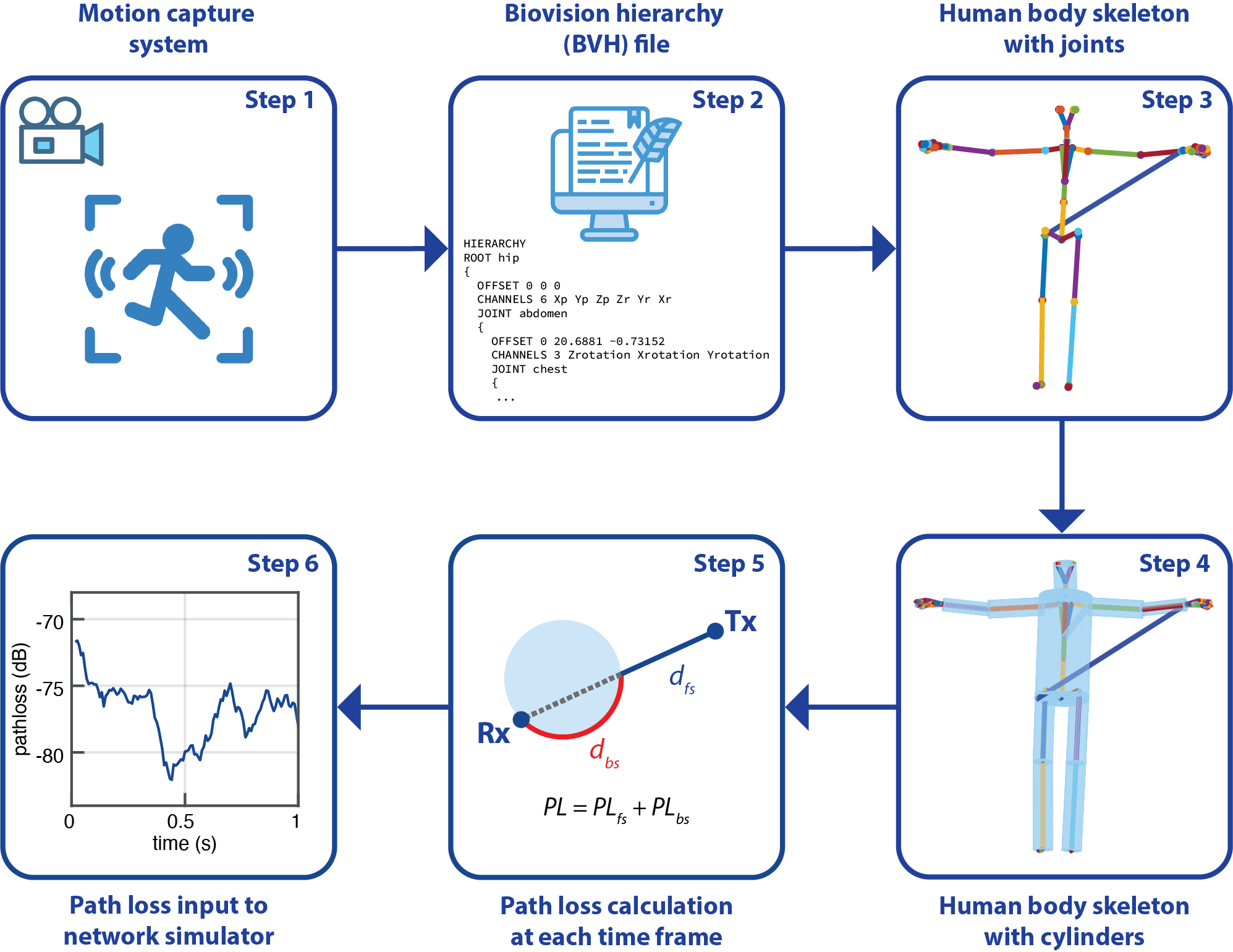}}
\caption{Generating a realistic human body path-loss model for use in WBAN simulators.}
\label{fig:bvh}
\end{figure*}

Fig.~\ref{fig:bvh} shows the steps that are taken to generate realistic path-loss values used as inputs to our network simulator. The first step consists of capturing the physical extent and location of the body while holding a posture or performing a specific activity, e.g.\ walking. This is recorded from human subjects wearing fixed markers on their body using a dedicated motion capture system. In Step 2, these results are saved in Biovision Hierarchy (BVH) file format~\cite{meredith2001motion} that includes the hierarchy of body joints and their positions at each point in time. We use BVH files from a dataset~\cite{mocap} recorded by a Vicon motion capture system consisting of 12 infrared MX-40 cameras with \SI{120}{\Hz} sampling rate and 41 markers worn on subjects.

Steps 3 to 6 are performed in MATLAB to extract the path-loss from BVH files. In Step 3, the skeleton hierarchy of the specific subject is imported and the connections between their joints are drawn. The model is then scaled to transform the skeleton to any arbitrary height and to make the model more generic. This is repeated for each time frame in the BVH files. Step 4 involves forming cylinders to approximate the body's volume. We used the height and cylinder dimensions of average adults in the US based on statistical data reported in~\cite{fryar2016anthropometric}.

Step 5 shows the transmitter (Tx), receiver (Rx), and torso from the top view. In order to calculate the path-loss while considering the shadowing effect of the torso, the Tx-Rx distance is divided into two portions: $d_{fs}$, the free space portion that does not intersect with the torso, and $d_{bs}$ accounting for the body surface portion. The latter is calculated as the shortest curvilinear distance between the two intersecting points on the cylinder surface (red helical path in Fig.~\ref{fig:bvh}). We use the Friis path-loss formula as a function of $d_{fs}$ (normalized to \SI{1}{\meter}) at \SI{2.4}{\giga\hertz} for the free space portion:

\begin{equation}
PL_{fs}[dB] = 20 \log_{10}(d_{fs}) + 40.0542
\label{eq:fspl}
\end{equation}

For the body surface part, we use the IEEE 802.15.6 CM3A model~\cite{yazdandoost2009channel}, where $N$ is a normal distribution with zero mean and $\sigma_N=3.8$, and $d_{bs}$ is normalized to \SI{1}{\meter}:

\begin{equation}
PL_{bs}[dB] = 6.6 \log_{10}(d_{bs}) + 36.1 + N
\label{eq:body}
\end{equation}

Finally, the total path-loss $PL$ at each time frame is computed as:

\begin{equation}
PL[dB] = PL_{fs}[dB] + PL_{bs}[dB]
\label{eq:pl}
\end{equation}

Step 6 illustrates the full path-loss after repeating these calculations for every frame. It is then saved in a file to be used as input to our full-stack network simulator~\cite{moin2017optimized} based on Castalia~\cite{boulis2007castalia}, which includes the estimated path-loss over time for a set of different on-body links observed in a WBAN.

\subsection{\SI{2.4}{\giga\hertz} Wearable Node test-bed}
\label{sec:testbed}

We utilized a custom-designed wearable node~\cite{moin2018emg} with a \SI{2.4}{\giga\hertz} radio chip (nRF51822, Nordic Semiconductor) to perform empirical measurements and test our proposed adaptive methods implemented on its ARM Cortex-M0 processor. The node also features a neural front-end~\cite{johnson2017implantable} which enables the acquisition of biosignals such as EMG and ECG.

\section{Adaptivity for Body Dynamic Channel}

In this section, we present the core idea of reusing sensor data that already exists in the WBAN to learn about the network state and adaptively reconfigure its parameters. The ultimate goal is to make the network more energy-efficient and robust with minimum added overhead. In the first part, we show that although the highly variable and often significant instantaneous path-loss around the human body makes communication challenging, the channel remains stable for hundreds of milliseconds during four typical daily activities. We take advantage of this property by learning the channel's behavior resulting from its kinematics and scheduling packet transmissions at appropriate moments. In the second part, we present an adaptive TPC method based on biosignals such as EMG and HR that are recorded in specific WBAN applications. These biosignals often correlate with body movements and postures, so the network can be adaptively reconfigured by monitoring them. Finally at the end of the section, we propose a state machine that integrates these two adaptive schemes for an energy-efficient and robust WBAN. Note that these are only two options we are currently considering with the goal of demonstrating the feasibility and efficacy of the adaptive approach; others may be added in the future.

\subsection{Human Body Kinematics}

\subsubsection{Channel stability analysis}

\begin{figure*}[!t]
\centerline{\includegraphics[width=0.77\textwidth]{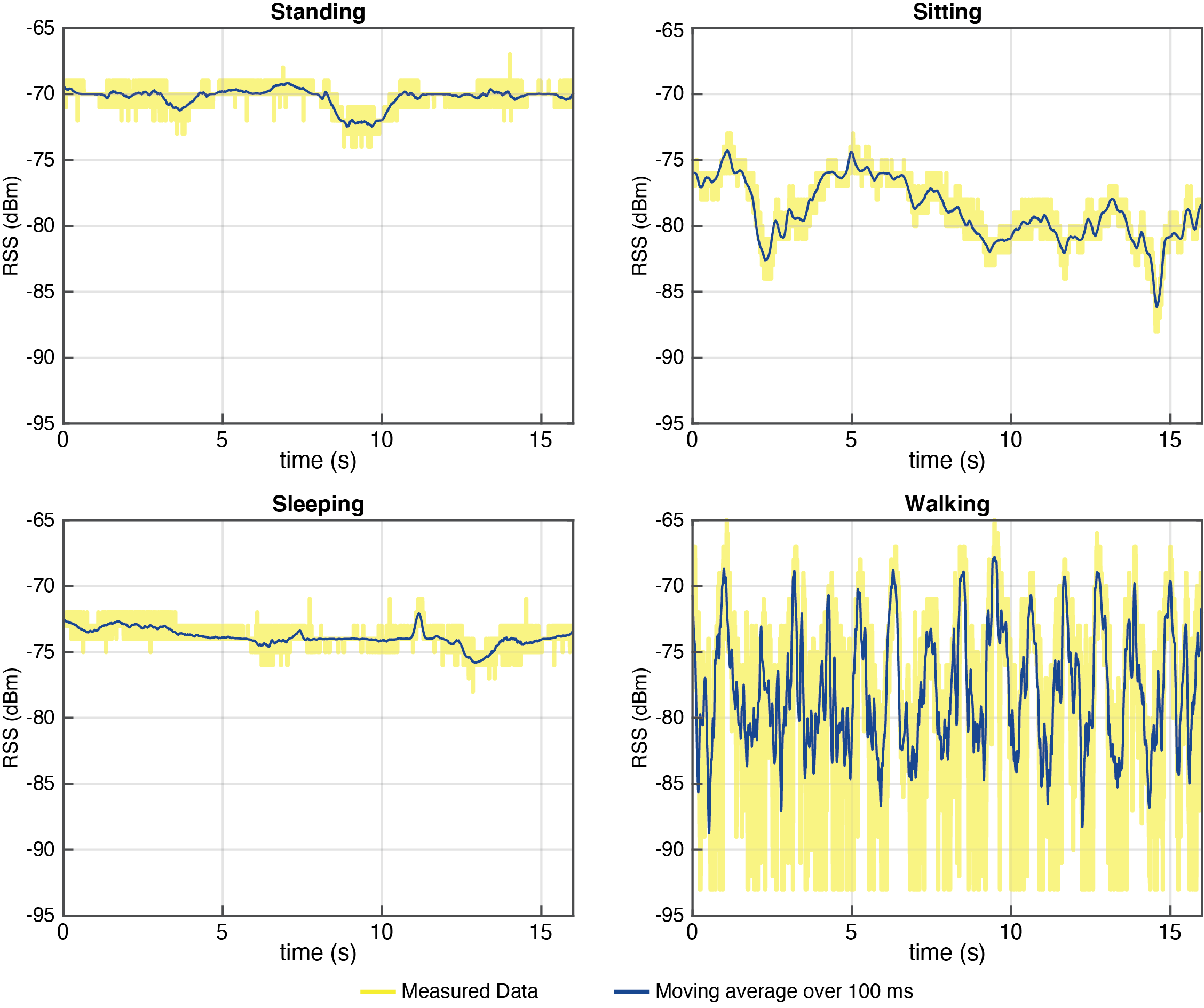}}
\caption{Received signal strength (RSS) measurements of the wireless channel between the left wrist and the right pants pocket while transmitting at \SI{0}{dBm} during four common daily activities. Yellow lines show the actual measured signal strength values with sample interval of \SI{1}{\milli\second} and blue traces show running average over \SI{100}{\milli\second} windows.}
\label{fig:activity}
\end{figure*}

Humans typically spend the majority of their daily lives in one of these states: standing, sitting, sleeping, or walking. Fig.~\ref{fig:activity} shows the path-loss variations and their moving average over \SI{100}{\milli\second} windows during these activities. We used our test-bed described in Section~\ref{sec:testbed} to perform these measurements with the transmitting node worn on the left wrist and the receiving node placed in the opposite side pants pocket. At a first glance, a relatively large fading (variation on signal strength) is observed for all activities, even for fairly passive ones such as sitting. However, this does not necessarily imply that the channel is unstable or unpredictable. For further analysis of channel stability, we computed two metrics for each activity: temporal autocorrelation and channel variation.

\begin{figure*}[!t]
\centerline{\includegraphics[width=0.8\textwidth]{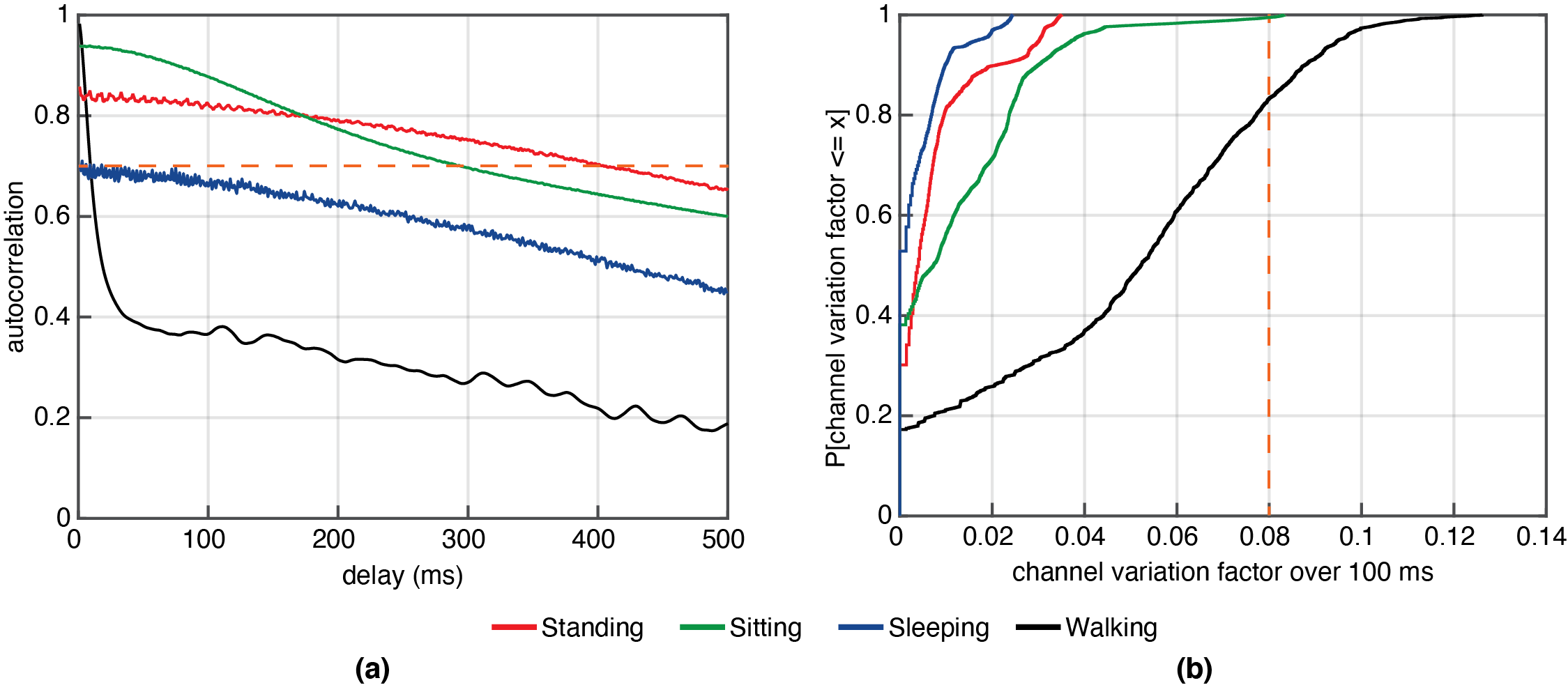}}
\caption{Analysis of wireless channel stability during the four activities shown in Fig.~\ref{fig:activity}. (a) Temporal autocorrelation as a function of delay time. (b) Cumulative distribution function for the channel variation factor calculated over periods of \SI{100}{\milli\second}.}
\label{fig:activity_corr}
\end{figure*}

Temporal autocorrelation~\cite{smith2009temporal} of wireless channels is often used to quantify stability in terms of channel coherence time, i.e.\ the period over which this correlation remains higher than $0.7$~\cite{smith2009temporal}. As Fig.~\ref{fig:activity_corr}(a) shows, the autocorrelation for the data gathered in Fig.~\ref{fig:activity} remains relatively high ($>0.7$) over several \SI{100}{\milli\second} for standing and sitting (static channels with similar environments). The channel coherence times are shorter for sleeping (different environment due to the bed) and walking (dynamic channel). The corresponding channel coherence times are \SI{404}{\milli\second}, \SI{294}{\milli\second}, \SI{18}{\milli\second}, and \SI{9}{\milli\second} for standing, sitting, sleeping, and walking, respectively.

Alternatively, it is possible to quantify stability using a channel variation factor (CVF), defined in~\cite{zhang2008stability} as:
\begin{equation}
CVF = \sqrt{\frac{\text{var}(\mathbf{h})}{\frac{1}{M}\Sigma_{m=0}^{M-1}h_m^2}}
\label{eq:cvf}
\end{equation}
where $\mathbf{h}=\{h_0,h_1,...,h_{M-1}\}$ is the channel impulse response in that period. In this case, the period over which a channel remains stable is defined as the period during which this quantity remains low ($<0.15$ for 90\% of measured samples~\cite{zhang2008stability}). We computed the CVF for periods of \SI{100}{\milli\second} and plotted its cumulative distribution function in Fig.~\ref{fig:activity_corr}(b) for all four activities. This metric shows that even in the case of walking in which channel variation is periodic, the CVF remains relatively low. The CVF remains below $0.08$ in 80\% of our observations and below $0.14$ in 100\% of our observations, which according to~\cite{zhang2008stability}, indicates a stable on-body channel. This relative stability of on-body channels enables accurate channel prediction across multiple communication frames, which can help to configure the transmit power, time and duration. The measurements shown in Figs.~\ref{fig:activity},\ref{fig:activity_corr} cover brief periods of time. Over a larger time-scale, the channel predictability is potentially higher due to recurrence of certain activities such as sleeping and transportation, and environments such as home and office.

\subsubsection{Periodic movements}
\label{sec:periodic}

Among the activities analyzed above, walking/running is the most challenging one with the fastest and most frequent changes in channel state. However, we can take advantage of its periodicity by predicting the peaks in received signal strength (RSS) and transmitting the packets only during those periods.

Inertial measurement units (IMUs) can be found in almost every WBAN~\cite{lecoutere2015wireless}, measuring a combination of linear acceleration and rotational rate. Therefore, the kinematics of the human body including the periodicity of walking would be reflected in the IMU reading. This signal has the same period as RSS and its peak times are correlated with RSS peaks.

\begin{figure*}[!t]
\centerline{\includegraphics[width=0.92\textwidth]{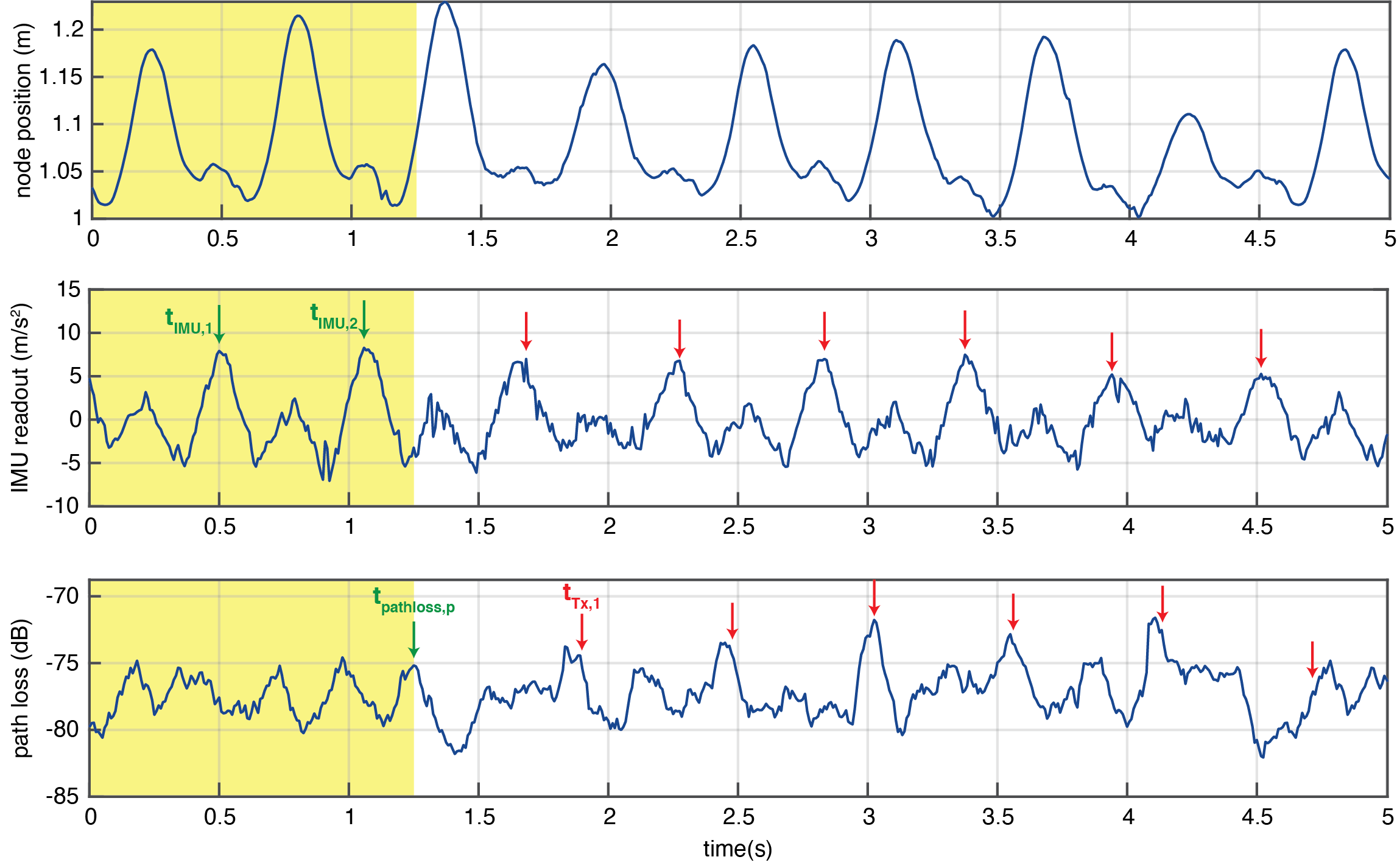}}
\caption{Left wrist position (top), IMU readouts (middle), and generated path-loss (bottom) for the wireless link between left wrist and opposite side pants pocket in walking scenario. The shaded portion indicates the initial calibration phase. The red arrows in the middle plot show detected peaks of IMU and in the bottom plot indicate the calculated packet transmission times.}
\label{fig:walking}
\end{figure*}

Fig.~\ref{fig:walking} illustrates a sample walking scenario generated by our channel emulator described in Section~\ref{sec:emulator}. The top plot shows the left wrist position in the camera coordinate system along the axis in which the person is walking (read from the BVH file). We then take its second derivative to generate what the IMU readout would look like (middle plot). Finally, the path-loss from the left wrist to the opposite side pants pocket calculated by our emulator is plotted at the bottom.

The process starts with a calibration phase (shaded area in Fig.~\ref{fig:walking}) during which we quantify the time difference between consecutive peaks in the IMU and find the $\alpha$ coefficient in~\eqref{eq:walking1} to link that to RSS peaks:

\begin{equation}
t_{IMU,2} + \alpha (t_{IMU,2} - t_{IMU,1}) = t_{pathloss,p}
\label{eq:walking1}
\end{equation}
where $t_{IMU,1}$ and $t_{IMU,2}$ are the first two peaks in the IMU readout and $t_{pathloss,p}$ is the first peak in RSS values that occurs after $t_{IMU,2}$. Note that RSS measurements are needed during the calibration phase that induces some overhead temporarily, but RSS values are no longer needed after the calibration is finished at $t_{pathloss,p}$. The next transmissions for $i>2$ will be scheduled at:

\begin{equation}
t_{Tx,i-2} = t_{IMU,i} + \alpha (t_{IMU,i} - t_{IMU,i-1})
\label{eq:walking2}
\end{equation}

This algorithm (see Algorithm~\ref{alg:periodic}) is computationally light for implementation on wearable nodes, unlike more complex operations such as dynamic time warping on sliding window of length $M$ and template of length $N$ with complexity $\mathcal{O}(MN)$ as used in~\cite{zang2017gait}. It is also inherently robust against gradual drifts in the peak periods since each transmission depends only on the previous two IMU peaks. However, a re-calibration is required in the rare cases of a significant change in walking pace or pattern, leading to a new $\alpha$ coefficient. These situations can be detected from the IMU readouts or multiple consecutive packet drops, triggering a re-calibration. Moreover, no additional control packet needs be communicated between receiver and transmitter after the calibration phase, since the packet transmission schedule depends only on IMU readouts. This significantly reduces the control overhead of most adaptive algorithms.

\begin{algorithm}[t]
\caption{Scheduling algorithm for periodic movements.}\label{alg:periodic}
\begin{algorithmic}[1]
\renewcommand{\algorithmicrequire}{\textbf{Input:}}
\renewcommand{\algorithmicensure}{\textbf{Output:}}
\Require
\Statex $IMU\_data$ \Comment{IMU readout}
\Statex $pathloss$ \Comment{needed only during calibration}
\Ensure
\Statex $\{t_{Tx,i}\}$ \Comment{scheduled transmission times}
\State $i \gets 1$
\While{receiving IMU readouts}
\State $t_{IMU,i} \gets \textsc{PeakDetection}(IMU\_data)$
\If {$i==2$} \Comment{calibration phase}
\State $t_{pathloss,p} \gets \textsc{PeakDetection}(pathloss)$
\State $\alpha \gets \textsc{RunCalib}(t_{IMU,1},t_{IMU,2},t_{pathloss,p})$
\ElsIf{$i>2$} \Comment{scheduled transmissions}
\State $t_{Tx,i-2} \gets t_{IMU,i} + \alpha (t_{IMU,i} - t_{IMU,i-1})$
\EndIf
\State $i \gets i+1$
\EndWhile
\end{algorithmic}
\end{algorithm}

\subsection{Electrical Biosignals}

One of the applications of a WBAN is health monitoring by hosting either generic biosensors such as body temperature and HR, or more specialized systems such as smart prostheses. The acquired biosignals provide indirect information on the network state. We propose here methods to reconfigure network parameters based on these biosignals.

\subsubsection{EMG}
\label{sec:emg}

Smart prostheses consist of a sensing system placed on the residual limb, a controller node responsible to make decisions based on the sensed signals, and an actuator node to perform the mechanical movements. Having a robust wireless link between these components is crucial since a failure in connectivity may lead to negative outcomes in the actuation, e.g.\ during a grasp task such as handling a cup of coffee. The same challenges of wireless channel around the human body exist here. Specifically, the interaction of the prosthesis with surrounding objects would most likely change the channel due to fading. We propose a learning-based adaptive algorithm that reconfigures the network when the prosthesis is in action.

Recording EMG using surface or implanted electrodes is one of the main input modalities to current prosthetic controllers~\cite{castellini2009surface,cipriani2008shared}. It enables the detection of patient's muscle contractions and interpretation of voluntary movement intentions. An increase in EMG activity implies subsequent movements of the prosthetic limb and consequently, changes in the wireless channel status are expected. 

\begin{figure*}[!t]
\centerline{\includegraphics[width=0.9\textwidth]{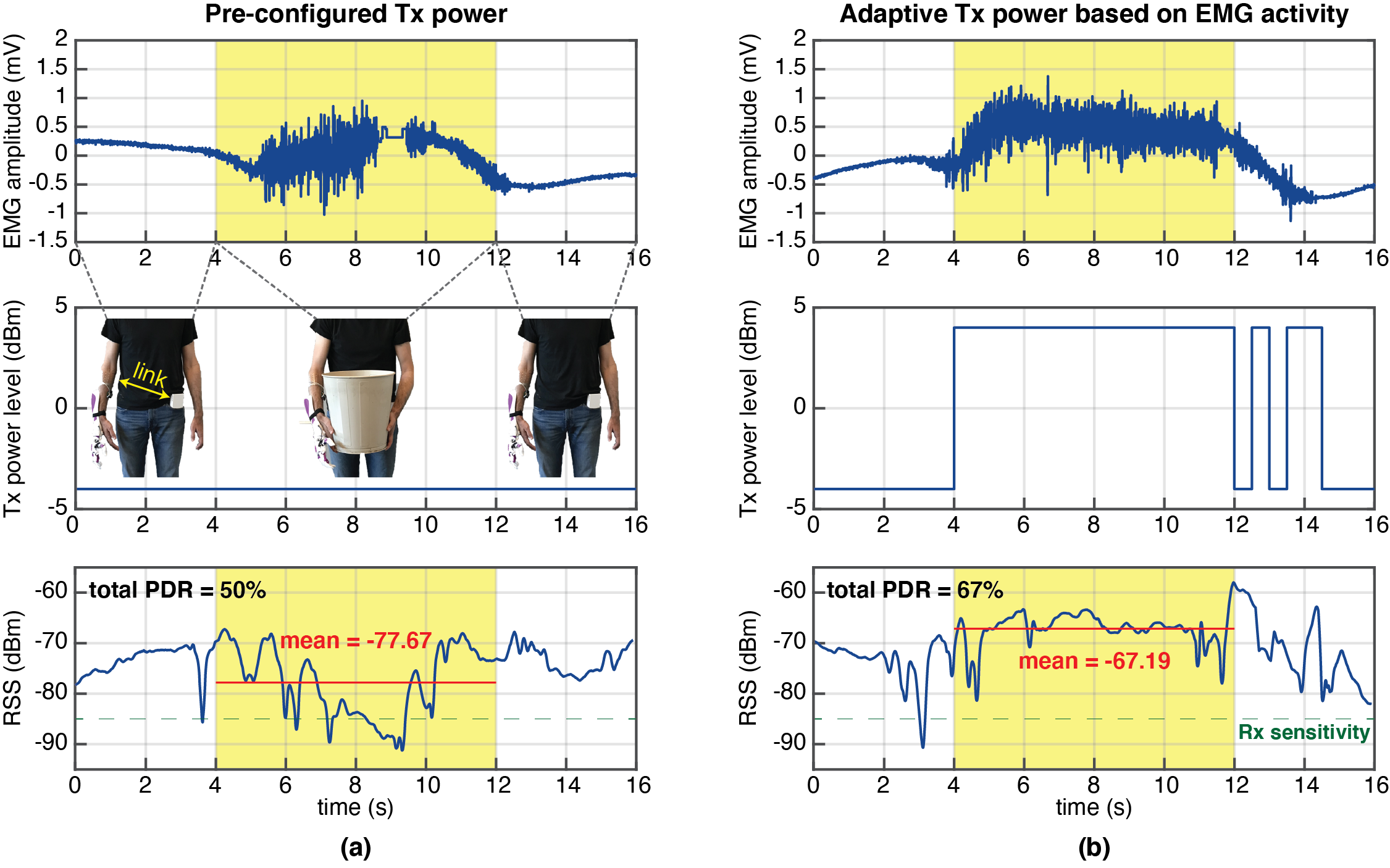}}
\caption{EMG-controlled TPC experimental results. The raw EMG signal (top), transmission power level (middle), and received signal strength (bottom) are plotted for: (a) a fixed transmission power scenario, and (b) an adaptive TPC based on EMG activity. Shaded areas correspond to holding the metal bucket.}
\label{fig:emg}
\end{figure*}

Fig.~\ref{fig:emg}(a) shows a scenario in which a prosthetic device attached to the right arm is used to hold a metal bucket. EMG from the forearm is recorded using the test-bed introduced in Section~\ref{sec:testbed}. The wireless link between the prosthetic node (containing both sensor and actuator) worn on the right forearm and the controller node placed in the opposite pants pocket is disrupted due to actuation movements and object shadowing effect. Our adaptive solution (see Algorithm~\ref{alg:emg}) uses a binary classifier to detect muscle activity and compensates for the degradation in wireless channel quality by readjusting the transmission power level. Note that the EMG signal is constantly measured for the control of the prosthetic arm anyways. Hence using that information for the management of the wireless network comes with zero overhead.

The radio transmission power is set to a low value of $P_{Tx,l}$ by default to keep the interference and power consumption low. The maximum ($V_{EMG,max}$) and minimum ($V_{EMG,min}$) values of EMG activity are calculated in windows of \SI{100}{\milli\second}. The node increases its transmission power to $P_{Tx,h}$ to guarantee wireless link quality under the following condition that occurs during the engagement of the prosthetic arm:

\begin{equation}
V_{EMG,max} - V_{EMG,min} > V_{EMG,thr},
\label{eq:emg}
\end{equation}
where $V_{EMG,thr}$ is preset by training data as a threshold on EMG voltage. As experimental results show later in Section~\ref{sec:results}, such a low overhead algorithm can considerably improve the link robustness. It should be noted that this is a very basic classification of recorded EMG signals with potential for significant improvements by using more complex algorithms~\cite{moin2018emg}.

\begin{algorithm}[!t]
\caption{EMG-controlled TPC algorithm.}\label{alg:emg}
\begin{algorithmic}[1]
\renewcommand{\algorithmicrequire}{\textbf{Input:}}
\renewcommand{\algorithmicensure}{\textbf{Output:}}
\Require
\Statex $V_{EMG}$ \Comment{EMG amplitude}
\Statex $V_{EMG,thr}$ \Comment{set based on training data}
\Ensure
\Statex $P_{Tx}$ \Comment{radio transmission power}
\State $V_{EMG,max} \gets \SI{0}{\milli\volt}$
\State $V_{EMG,min} \gets \SI{100}{\milli\volt}$
\State $i \gets 1$ \Comment{samples counter every \SI{1}{\milli\second}}
\State $P_{Tx} \gets P_{Tx,l}$
\While{receiving EMG samples}
\If {$V_{EMG}>V_{EMG,max}$}
\State $V_{EMG,max} \gets V_{EMG}$
\ElsIf{$V_{EMG}<V_{EMG,min}$}
\State $V_{EMG,min} \gets V_{EMG}$
\EndIf
\If {$i==100$} \Comment{reached end of \SI{100}{\milli\second} period}
\If {$V_{EMG,max} - V_{EMG,min} > V_{EMG,thr}$}
\State $P_{Tx} \gets P_{Tx,h}$
\Else
\State $P_{Tx} \gets P_{Tx,l}$
\EndIf
\State $V_{EMG,max} \gets \SI{0}{\milli\volt}$
\State $V_{EMG,min} \gets \SI{100}{\milli\volt}$
\State $i \gets 0$
\EndIf
\State $i \gets i+1$
\EndWhile
\end{algorithmic}
\end{algorithm}

\subsubsection{Heart rate}
\label{sec:ecg}

Heart rate (HR) monitoring is one of the most common applications of WBAN. It is achieved by optical or electrical measurements using nodes worn typically on a wrist or the chest. This signal varies according to the human activity level. At the same time, an increase in physical activity often leads to degradation in wireless channel quality. This degradation can be compensated by adaptively reconfiguring the network as explained in Algorithm~\ref{alg:ecg}. Observe that an increase in physical activity may be more accurately detected by an IMU, given that an increase in HR may also be caused by static stress conditions a subject is experiencing. However, the HR-based TPC method matches well the simple HR monitoring patches that are not necessarily equipped with an IMU.

\begin{figure*}[!t]
\centerline{\includegraphics[width=\textwidth]{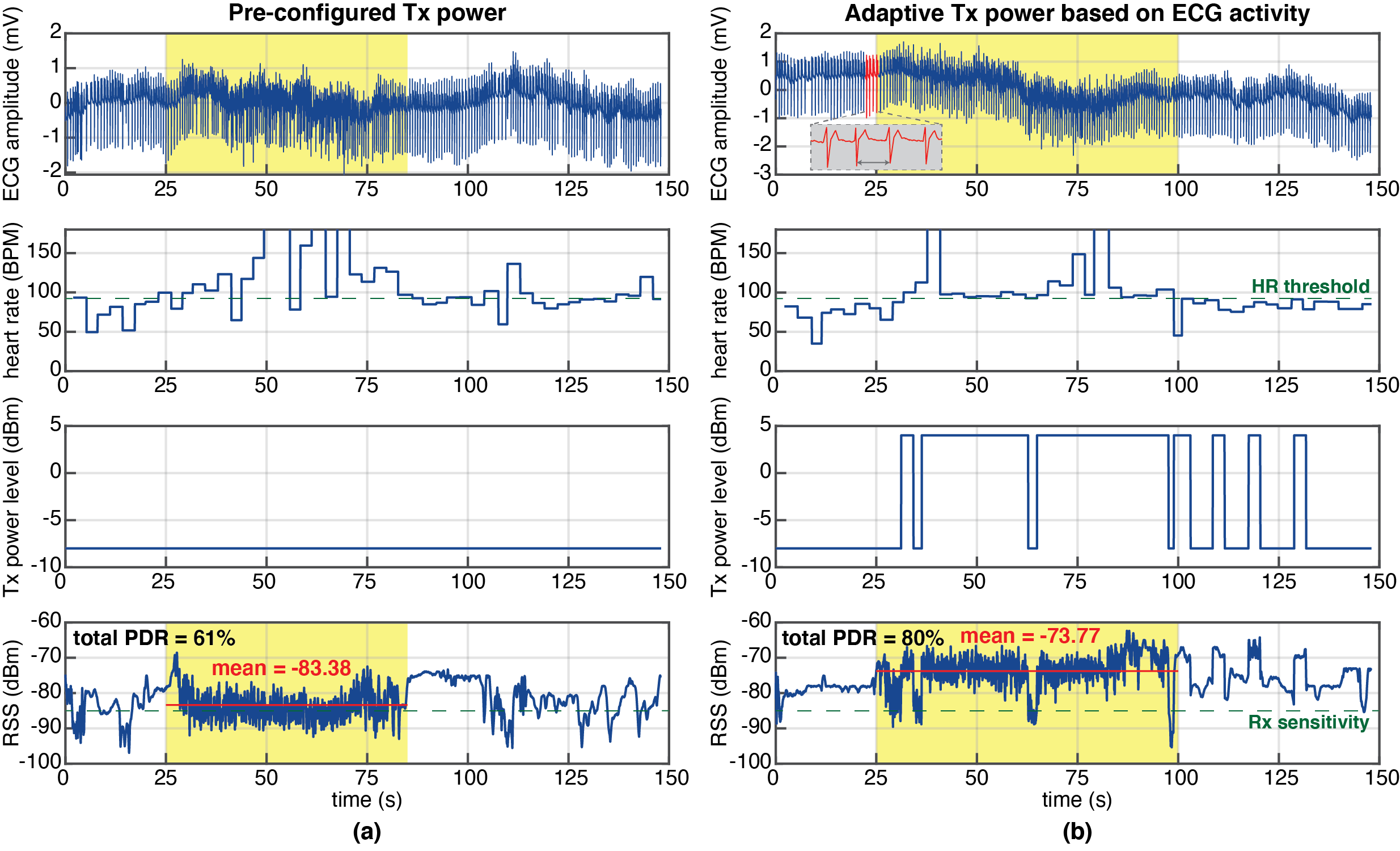}}
\caption{HR-controlled TPC experimental results. The raw ECG signal (first row), extracted HR (second row), transmission power level (third row), and received signal strength (last row) are plotted for: (a) a fixed transmission power scenario, and (b) an adaptive TPC based on HR. Shaded and unshaded areas correspond to increased physical activity level during walking and low activity during sitting down, respectively.}
\label{fig:ecg}
\end{figure*}

Fig.~\ref{fig:ecg}(a) shows ECG signal recorded from chest using our test-bed explained in Section~\ref{sec:testbed}. A simple algorithm, denoted as \textsc{CalcHR} in Algorithm~\ref{alg:ecg}, was developed to extract the HR from ECG recordings (Fig.~\ref{fig:ecg}, second row), and implemented on the transceiver processor. The function keeps track of the ECG maximum ($V_{ECG,max}$) and minimum ($V_{ECG,min}$), triggers a counter whenever $V_{ECG}$ passes $V_{ECG,thr}$:

\begin{equation}
V_{ECG,thr} = V_{ECG,min} + 0.25*(V_{ECG,max}-V_{ECG,min}),
\label{eq:ecg}
\end{equation}
and reads the counter value which corresponds to the interval between two consecutive ECG peaks, i.e.\ the HR period. Performing only addition/subtraction, bit-shifting (multiply by $0.25$) and comparisons makes it computationally light with negligible effect on the energy consumption. Note that the occasional spikes in HR are due to undetected heart beats in our simple algorithm (which do not impact the efficacy of the overall control scheme). The radio transmission power is kept at a low level of $P_{Tx,l}$ during body static state by default. When the HR goes above a threshold value ($HR_{thr}$) due to physical activity, the radio automatically increases the transmission power to $P_{Tx,h}$ to compensate for channel fading and variations. This improves the link quality as experimental results show later in Section~\ref{sec:results}.

\begin{algorithm}[!t]
\caption{HR-controlled TPC algorithm.}\label{alg:ecg}
\begin{algorithmic}[1]
\renewcommand{\algorithmicrequire}{\textbf{Input:}}
\renewcommand{\algorithmicensure}{\textbf{Output:}}
\Require
\Statex $V_{ECG}$ \Comment{ECG amplitude}
\Statex $HR_{thr}$ \Comment{set based on training data}
\Ensure
\Statex $P_{Tx}$ \Comment{radio transmission power}
\State $i \gets 1$ \Comment{samples counter every \SI{1}{\milli\second}}
\State $P_{Tx} \gets P_{Tx,l}$
\While{receiving ECG samples}
\State $HR \gets \textsc{CalcHR}(V_{ECG})$ \Comment{calculate HR}
\If {$i==3000$} \Comment{update $P_{Tx}$ every \SI{3}{\second}}
\If {$HR>HR_{thr}$}
\State $P_{Tx} \gets P_{Tx,h}$
\Else
\State $P_{Tx} \gets P_{Tx,l}$
\EndIf
\State $i \gets 0$
\EndIf
\State $i \gets i+1$
\EndWhile
\end{algorithmic}
\end{algorithm}

\subsection{Integrated State Machine}

\begin{figure}[!t]
\centerline{\includegraphics[width=0.8\columnwidth]{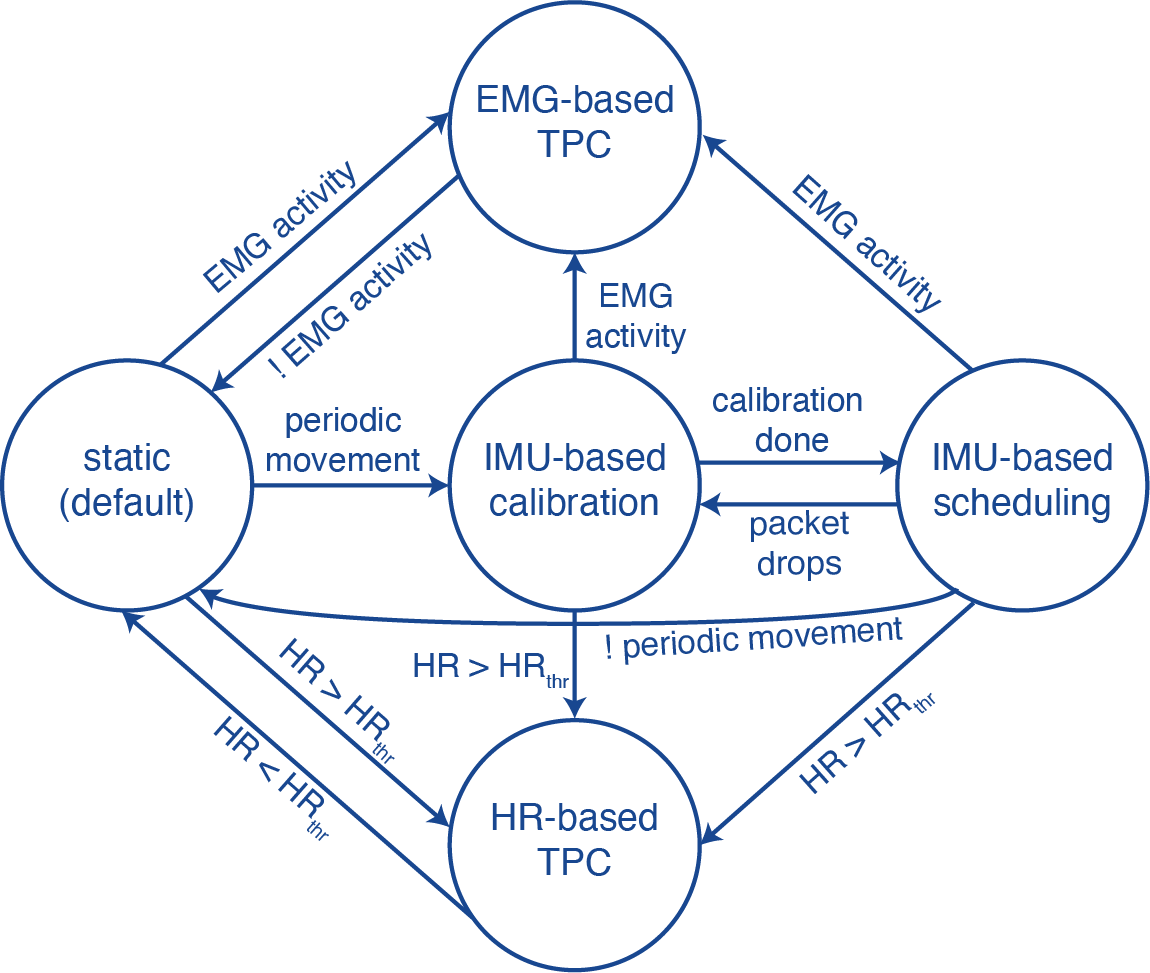}}
\caption{The state transition diagram of our adaptive WBAN.}
\label{fig:state}
\end{figure}

The adaptive schemes introduced in this section can all be integrated into a WBAN stack as shown in the state diagram of Fig.~\ref{fig:state}. The network stays in \textit{static} state by default that corresponds to activities such as sitting and sleeping. When the IMU detects a periodic movement such as walking, \textit{IMU-based calibration} is triggered and the periodic pattern of movement is learned as described in Section~\ref{sec:periodic}. The network transmits packets based on IMU readouts as long as the state remains unchanged. A re-calibration might be necessary if multiple packet drops happen due to significant changes in the walking pattern or pace. If EMG activity in the forearm is detected at any point in time, the network transitions to \textit{EMG-based TPC} (Section~\ref{sec:emg}). Similarly, if the HR goes above its threshold, \textit{HR-based TPC} (Section~\ref{sec:ecg}) will be engaged. Note that IMU-based activity recognition for classifying static states (e.g.\ sitting vs.\ sleeping) can be done~\cite{khan2010triaxial} to further optimize the network configurations such as routing tables and MAC layer parameters.

\section{Experimental Results and Performance Evaluation}
\label{sec:results}

In this section, we present experimental results of the network energy efficiency and robustness after implementing our proposed adaptive scheme.

\subsection{Periodic Movements}

\begin{figure}[!t]
\centerline{\includegraphics[width=0.6\columnwidth]{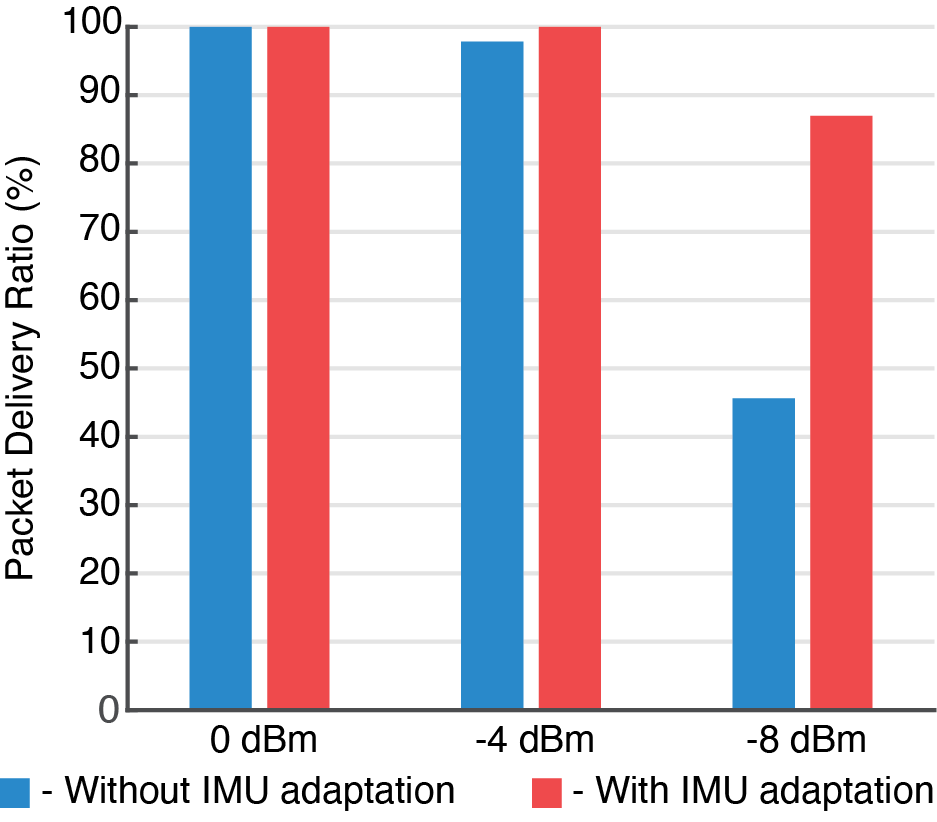}}
\caption{Packet delivery ratio for different transmission power levels during two minutes of walking with and without IMU-based adaptation.}
\label{fig:walking_results}
\end{figure}

We imported the emulated path-loss from Section~\ref{sec:periodic} into our network simulator~\cite{moin2017optimized}. The application layer was programmed to transmit packets in \SI{100}{\milli\second} intervals from the left wrist to the right pants pocket, mimicking a smart watch to smart phone communication scenario. The physical layer parameters are based on a Bluetooth chipset (nRF51822, Nordic Semiconductor) with \SI{-85}{dBm} receiver sensitivity in \SI{2}{Mbps} data rate mode. The blue bars in Fig.~\ref{fig:walking_results} show packet delivery ratio (PDR) for different radio transmission power levels during two minutes of walking. While PDR can reach 100\% by operating in a high-power mode of \SI{0}{dBm} transmission power and still remains reliable in \SI{-4}{dBm} mode, it will go below 50\% after decreasing the transmission power level to \SI{-8}{dBm}.

We then repeated the same simulations using our proposed adaptive algorithm based on kinematics (Section~\ref{sec:periodic}). The application layer generated packets at the same rate, but packets were buffered to be transmitted at RSS predicted peaks (red arrows in Fig.~\ref{fig:walking}, bottom plot). The red bars in Fig.~\ref{fig:walking_results} show that using our adaptive scheduling, PDR remains at 100\% after reducing transmission power level to \SI{-4}{dBm}, and shows significant improvement after going down to \SI{-8}{dBm} compared to the case without IMU (87\% versus 46\%). The reduced transmission power has the advantage of lower power consumption in wearable resource-limited nodes as well as less interference with other WBAN nodes.

\subsection{EMG- and HR-Controlled TPC}

Fig.~\ref{fig:emg} illustrates the experimental results of a sample trial in the prosthetic arm scenario explained in Section~\ref{sec:emg}. An increase in EMG amplitude can be observed in the top plot of Fig.~\ref{fig:emg}(a) between \SI{4}{\second} and \SI{12}{\second} when the subject transfers from rest state to holding the metal bucket. The bottom plot shows RSS when transmitting packets with fixed transmission power of $P_{Tx,l}=$\SI{-4}{dBm}. The RSS is degraded during the hold task (shaded area of plots) which led to a total PDR of 50\%.

Fig.~\ref{fig:emg}(b) shows the results for the same experiment after implementing our adaptive algorithm on the radio processor with $V_{EMG,thr}=$\SI{610}{\micro\volt}. The transmission power level is automatically boosted to $P_{Tx,h}=$\SI[retain-explicit-plus]{+4}{dBm} upon detection of EMG activity to assure higher link reliability. Consequently, the mean RSS is increased more than \SI{10}{dBm} during the prosthetic arm engagement and total PDR is increased to 67\%. Note that around $t=$\SI{9}{\second} in Fig.~\ref{fig:emg}(a), a brief period of packet drops is observed due to the channel fading. This is avoided by running the EMG-based TPC algorithm at the sensing node in Fig.~\ref{fig:emg}(b).

\begin{figure}[!t]
\centerline{\includegraphics[width=0.8\columnwidth]{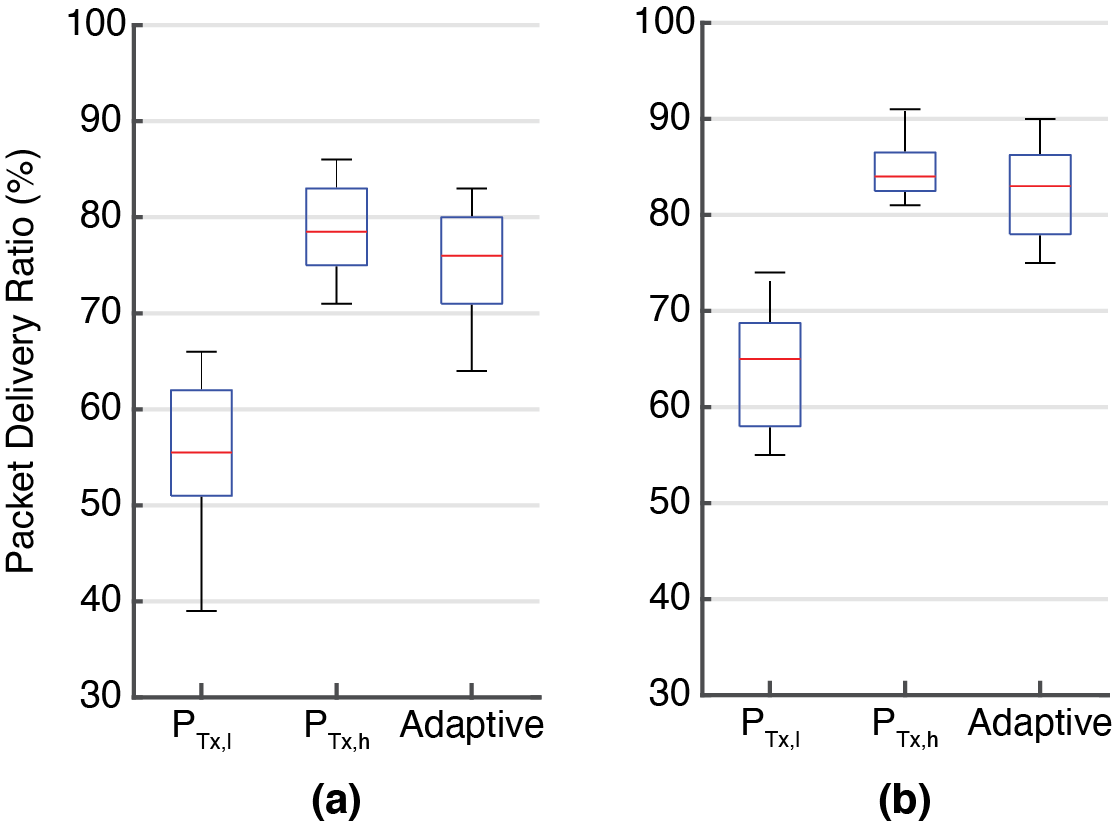}}
\caption{Packet delivery ratio results for fixed and adaptive transmission power levels: (a) EMG-controlled TPC for $P_{Tx,l}=$\SI{-4}{dBm} and $P_{Tx,h}=$\SI[retain-explicit-plus]{+4}{dBm} ($N=10$ trials). (b) HR-controlled TPC for $P_{Tx,l}=$\SI{-8}{dBm} and $P_{Tx,h}=$\SI[retain-explicit-plus]{+4}{dBm} ($N=5$ trials).}
\label{fig:exg_results}
\end{figure}

We repeated these experiments for $N=10$ trials using both our adaptive method and pre-configured transmission powers at $P_{Tx,l}=$\SI{-4}{dBm} and $P_{Tx,h}=$\SI[retain-explicit-plus]{+4}{dBm}. As the boxplots in Fig.~\ref{fig:exg_results}(a) show, our EMG-controlled TPC method has 20\% higher PDR than the fixed $P_{Tx,l}$ case, and only 4\% lower than the constant transmission at $P_{Tx,h}$.

Fig.~\ref{fig:ecg} presents the experimental results of a sample trial in the HR-controlled TPC scenario explained in Section~\ref{sec:ecg}. The subject is wearing the test-bed node on the left wrist and the receiver node is placed in the right pants pocket. The experiment starts with rest in sitting state followed by an increase in activity (walking with high pace) and back to rest. The bottom plot in Fig.~\ref{fig:ecg}(a) shows RSS when transmitting packets with fixed transmission power of $P_{Tx,l}=$\SI{-8}{dBm}. The RSS is degraded during walking (shaded area of plots) which led to a total PDR of 61\%.

The neural front-end of our test-bed is connected to an Ag/Ag-Cl electrode placed on the subject's chest to measure their ECG. The implemented algorithm on the radio processor extracts the HR by measuring the distance between ECG peaks and adaptively toggles the transmission power level to $P_{Tx,h}=$\SI[retain-explicit-plus]{+4}{dBm} when the HR goes higher than $HR_{thr}=$\SI{92}{bps} threshold. Fig.~\ref{fig:ecg}(b) shows that our proposed adaptive scheme increases the mean RSS during body movements by \SI{9.6}{dBm}, causing the total PDR to reach 80\% in this trial. After repeating these experiments for $N=5$ times (Fig.~\ref{fig:exg_results}(b)), we observed an average PDR increase of 18\% using the adaptive HR-based TPC compared to the fixed $P_{Tx,l}$ case, and only 2\% decrease from the constant transmission case at $P_{Tx,h}$.

\subsection{Power Savings in Sample Scenarios}

In order to show the energy efficiency of our adaptive method, we performed a comparison between scenarios with and without engaging our proposed algorithm. The parameters used for this comparison are listed in Table~\ref{tab:power}. We assumed failed packets would be successfully delivered after one re-transmission attempt. 

In the case of walking, a 27\% reduction in power consumption was achieved by using the IMU-based scheduling. For the prosthetic arm scenario, constantly transmitting at $P_{Tx,h}$ to have a reliable link, caused a 31\% increase in power consumption compared to the fixed $P_{Tx,l}$ case. Incorporating the adaptive EMG-controlled TPC, however, led to only a 16\% increase in power consumption while the link reliability was preserved. From a power saving perspective, our adaptive EMG-controlled TPC decreased the power consumption by 12\% compared to the fixed $P_{Tx,h}$ case, while preserving the PDR (only 4\% loss). Note that the overhead power of running our adaptive scheme is negligible since the IMU data and biosignals are already available in the network and the algorithms are computationally light.

\begin{table}[!t]
\caption{Parameters used in sample scenarios.}
\label{tab:power}
\centering
\begin{tabular}{|c|c|} \hline
$P_{Tx}$ for Walking & \SI{-8}{dBm}\\ \hline
$P_{Tx,l}$ for Prosthesis & \SI{-4}{dBm}\\ \hline
$P_{Tx,h}$ for Prosthesis & \SI[retain-explicit-plus]{+4}{dBm}\\ \hline
Power Consumption at \SI{-8}{dBm} & \SI{21}{\milli\watt}\\ \hline
Power Consumption at \SI{-4}{dBm} & \SI{24}{\milli\watt}\\ \hline
Power Consumption at \SI[retain-explicit-plus]{+4}{dBm} & \SI{31.5}{\milli\watt}\\ \hline
\end{tabular}
\end{table}

\section{Conclusion}

In this paper, we demonstrated the efficacy of an adaptive WBAN approach that reconfigures the network based on body kinematics and biosignals. By careful scheduling packet transmissions derived from the IMU data, we gained a 41\% increase in the PDR while keeping the transmission power low to reduce interference and power consumption. Additionally, the experimental results showed up to 20\% boost in the average PDR when engaging our proposed EMG- and HR-controlled TPC methods. The presented body dynamic channel emulator is of value in other studies that require a realistic channel model, especially since it builds on the extensive BVH datasets representing a broad range of more complex physical activities such as jumping and dancing. Future opportunities include adding more biosignals and sensors to the proposed adaptive scheme, possibly increasing the network energy-efficiency and robustness under various scenarios. The ultimate goal is to have these schemes added as a vertical control plane to existing protocols such as the IEEE 802.15.6 standard, enabling them to learn, identify and switch between network states as shown in Fig.~\ref{fig:state}.

\section*{Acknowledgments}

The authors would like to thank Andy Zhou, George Alexandrov, Prof.\ Elad Alon and Prof.\ Adam Wolisz for their feedback. The motion capture data used in this project was obtained from \url{mocap.cs.cmu.edu} database funded by NSF EIA-0196217.

\bibliographystyle{IEEEtran}
\bibliography{IEEEabrv,references.bib}

\end{document}